\documentclass[aps,superscriptaddress,amsfonts,amssymb,amsmath,showpacs,nofootinbib, twocolumn]{revtex4-1}
\usepackage[dvips]{graphicx}

\begin{document}

\title{Slowly rotating neutron stars with small differential rotation: equilibrium models and oscillations in the Cowling approximation}

\author{Cecilia Chirenti} 
\email{e-mail: cecilia.chirenti@ufabc.edu.br} 
\affiliation{Centro de Matem\'atica, Computa\c c\~ao e Cogni\c c\~ao, UFABC, 09210-170 Santo Andr\'e, SP, Brazil}
\author{Jozef Sk\'akala}
\email{e-mail: jozef.skakala@ufabc.edu.br}
\affiliation{Centro de Matem\'atica, Computa\c c\~ao e Cogni\c c\~ao, UFABC, 09210-170 Santo Andr\'e, SP, Brazil}
\author{Shin'ichirou Yoshida}
\email{e-mail: yoshida@ea.c.u-tokyo.ac.jp}
\affiliation{
Department of Earth Science and Astronomy, Graduate School of Arts and Sciences, University of Tokyo, Komaba, Meguro-ku
3-8-1, 153-8902 Tokyo, Japan}

\begin{abstract}
Newly born neutron stars can present differential rotation, even if later
it should be suppressed by viscosity or a sufficiently strong magnetic field. And in this early stage of its life, a
neutron star is expected to have a strong emission of gravitational waves, which could be influenced by the differential rotation. We present here a new formalism for modelling differentially rotating neutron stars: working on the slow rotation approximation and assuming a small degree of differential rotation, we show that it is possible to separate variables in the Einstein field equations. The dragging of inertial frames is determined by solving three decoupled ODEs. After we establish our equilibrium model,
we explore the influence of the differential rotation on the f and r-modes of oscillation of the neutron star in the Cowling approximation, and we also analyze an effect of the differential rotation on the emission of gravitational radiation from the f-modes. We see that the gravitational radiation from the f-modes is slightly suppressed by introducing differential rotation to the equilibrium stars.
\end{abstract}

\maketitle

\section{Introduction}

Differential rotation, until it becomes suppressed by viscosity or strong enough magnetic fields \cite{Shapiro1,Shapiro2}, might play an important role in the evolution of a newly born neutron star. (For a typical neutron star it takes between 10-100 years to become uniformly rotating \cite{Yoshida1}.) The equilibrium stellar models representing neutron star's differential rotation were explored in some older papers \cite{Hartle1, Will}, and the oscillation frequencies for some types of fluid modes were calculated later in \cite{Yoshida1, Yoshida2, Passamonti1, Passamonti2}.  

 In this work we explore the evolution of linear perturbations in a slowly rotating neutron star with a polytropic equation of state. Moreover, the rotation profile of the star represents a first order deviation from the uniform rotation. We generalize the old semi-analytical results of Hartle for the uniformly rotating equilibrium model \cite{Hartle2} to a small deviation from the uniform rotation following the relativistic j-constant law. In particular we show that under a first order deviation from the slow uniform rotation one can still separate spherical harmonics and obtain only a very small number of non-zero terms in their expansion. Furthermore, similar to the uniformly rotating case \cite{Hartle2}, one can find an exact analytic solution for the metric dragging function outside the star. 

In Section \ref{sec:model} of this paper we use the consistency conditions imposed by the equilibrium model to constrain the value of the parameter representing the differential rotation, (for the given equilibrium parameters of the star). Such a relatively simple equilibrium model with the constrained value of the differential rotation parameter is then used in Section \ref{sec:perturb}, in the Cowling approximation, to explore the various types of fluid modes. To obtain numerical results for the modes, in Section \ref{sec:numerics} we evolve the initial perturbation in time using a 2D Lax-Wendroff scheme \cite{Mitchell}.
Our choice of a time evolution treatment is motivated by the co-rotation problem that affects the usual eigenvalue approach for obtaining the mode frequencies.
The f-modes were computed for the slow rotation case in \cite{Passamonti1, Passamonti2} and we compare our results with the results from those papers, further constraining the domain of validity of the differential rotation parameter. In addition to the results of \cite{Passamonti1, Passamonti2}, we explore in more detail the behavior of the f-modes as a function of small values of the differential rotation parameter and we also find numerical values for some of the r-mode frequencies. (As can be seen in \cite{Rezzolla-r-modes} the r-modes excite differential rotation in the stellar fluid.)
Also, the f-mode eigenfunctions are extracted by using a pointwise discrete Fourier transform (DFT) on the evolution data. As the equilibrium stars have slow rotation with a low degree of differential rotation, the eigenmodes extracted show small change from their non-rotating counterpart. 
In Section \ref{sec:gw}, by using a quadrupole estimate of gravitational emission timescale, we see a tendency that the differential rotation slightly suppresses the gravitational emission. These results are consistent with the results of \cite{Kruger} for the rapidly rotating stars. Finally, we finish with our conclusions in Section \ref{sec:conclusions}.

\section{Equillibrium stellar model}
\label{sec:model}

Consider the background spacetime of a slowly rotating star:
\begin{equation*}
ds^2 = -e^{\nu} dt^{2}+e^{\lambda}
dr^{2}+r^{2}d\theta^{2}+r^{2}\sin^{2}(\theta)\left[d\phi-\omega
dt\right]^{2},
\end{equation*}
where $\nu$ and $\lambda$ are functions of $r$ and $\omega = \omega(r,\theta)$ is the frame dragging function. We will use a polytropic equation of state
$p = K \epsilon^{1 + 1/N}$,
where $p$ is the pressure and $\epsilon$ is the rest mass energy density of the star. The fluid rotation is described by the 4-velocity 
\[u^{(t,r,\theta,\phi)}=(e^{-\nu/2},~0,~0,~\Omega \cdot e^{-\nu/2}).\]
Further, we consider that the rotation of the fluid, $\Omega$, obeys the $j$-constant law:
\begin{equation}\label{j-const}
\Omega=\frac{\Omega_{c}+\gamma\cdot r^{2}\sin^{2}(\theta)
e^{-\nu}\omega}{1+\gamma\cdot r^{2}\sin^{2}(\theta) e^{-\nu}}~.
\end{equation}
The $\gamma$ parameter in the equation \eqref{j-const} describes the level of the differential rotation of the star.
Then the TOV equations remain unchanged under the slow rotation (in the linear order in $\Omega$), and the frame dragging parameter $\omega$ has to be a solution of the equation:
\begin{widetext}
\begin{eqnarray}\label{Hartle}
\frac{1}{r^{4}}\frac{\partial}{\partial r}\left[r^{4} e^{-(\nu+\lambda)/2} \cdot\frac{\partial \omega}{\partial r}\right]+\frac{ e^{(\lambda-\nu)/2}}{r^{2}\sin^{3}(\theta)} \frac{\partial}{\partial \theta}\left[\sin^{3}(\theta) \frac{\partial \omega}{\partial \theta}\right] -
16\pi e^{(\lambda-\nu)/2} \cdot (\epsilon+p)\left[\omega -
\frac{\Omega_{c}+\gamma\cdot r^{2}\sin^{2}(\theta)
e^{-\nu}\omega}{1+\gamma\cdot r^{2}\sin^{2}(\theta) e^{-\nu}}\right]
= 0.~~~~~
\end{eqnarray}
\end{widetext}

Furthermore, take the differential rotation to be also small, representing only a linear order perturbation from the uniform rotation case: $0<\gamma<<1$. Then one can expand $\omega$ in the $\gamma$ parameter as:
\begin{equation}
\omega(r,\theta)=\omega_{0}(r,\theta)+\gamma\cdot\omega_{1}(r,\theta)+O(\gamma^{2}).
\end{equation}
Here $\omega_{0}$ corresponds to the case of uniform rotation and as we know from \cite{Hartle2}, it depends only on $r$. (In the zeroth order expansion of \eqref{Hartle}  in $\gamma$ we obtain Hartle's equation for the uniform rotation problem \cite{Hartle2}.) Now, let us write the first order equation  (in $\gamma$), which represents a correction due to a small differential rotation modifying the uniform rotation problem. This will be:
\begin{widetext}
\begin{eqnarray}\label{Hartle-mod}
\frac{1}{r^{4}}\frac{\partial}{\partial r}\left[r^{4} e^{-(\nu+\lambda)/2}\cdot \frac{\partial \omega_{1}}{\partial r}\right]+\frac{e^{(\lambda-\nu)/2}}{r^{2}\sin^{3}(\theta)} \frac{\partial}{\partial \theta}\left[\sin^{3}(\theta) \frac{\partial \omega_{1}}{\partial \theta}\right] -
16\pi  e^{(\lambda-\nu)/2}\cdot (\epsilon+p)\left[\omega_{1} - (\omega_{0}-\Omega_{c}) r^{2}\sin^{2}(\theta) e^{-\nu}\right] = 0.~~~~~~~~
\end{eqnarray}
\end{widetext}
Due to the fact that $\omega_{0}$ does not depend on $\theta$, one can simplify the problem by decomposing the terms in the equation \eqref{Hartle-mod} into the vector spherical harmonics and one obtains (symbol  ``~$'$~'' means $r$-derivative):
\begin{widetext}
\begin{eqnarray}\label{decomposed}
[r^{4}j~\omega'_{1 \ell}]'=e^{\lambda}~j~r^{2}[\{\ell(\ell+1)-2\}\omega_{1\ell}+16\pi r^{2}(\epsilon+p)\{\omega_{1\ell}-r^{2}e^{-\nu} C_{\ell} [\omega_{0}-\Omega_{c}] \} ].~~~~~~~~
\end{eqnarray}
\end{widetext}
Here $j(r) = e^{-(\nu+\lambda)/2}$ and $C_{\ell}$ is a $\ell -th$ coefficient of the decomposition of $\sin^{2}(\theta)$ into vector spherical harmonics.
One can express the decomposition as
\begin{equation}
\sin^{2}(\theta)=\frac{4}{5}-\frac{2}{15}\left(\frac{15}{2}\cos^{2}(\theta)-\frac{3}{2}\right),
\end{equation}
and thus the only two non-zero coefficients $C_{\ell}$ of the decomposintion are ~$C_{1}=4/5$ ~and ~ $C_{3}=-2/15$.

The two linearly independent solutions of the equation \eqref{decomposed} behave for $\ell>1$, both close to zero and at infinity, as
\begin{equation}\label{asymp}
C_{+} r^{-(2+\ell)} + C_{-} r^{\ell-1}.
\end{equation}
The $\ell=1$ case shows the same behavior \eqref{asymp} at the infinity, but close to zero one has to be more careful: One can try to Taylor expand the solutions at the origin, proving that only one of the solutions is analytic around zero. Then one can naturally expect, that also in the case $\ell=1$ one of the solutions is singular at zero. (For more details see \ref{regularity}.)

It can be easily shown that in case $C_{\ell}=0$, ~$\ell>1$, the regular behavior of the solution at the infinity cannot be matched with the regular behavior at zero (see again appendix \ref{regularity}). This means for $C_{\ell}=0$ \emph{no} non-trivial relevant solutions exist. On the other hand, for $C_{\ell}\neq 0$ we can, through the Green function, construct everywhere regular non-trivial solutions. 
This means the frame dragging function $\omega$ can be expressed in the linear order of $\gamma$ as:
\begin{eqnarray}
\label{dragg}
\omega=\omega_{0}(r)+\gamma\cdot[~\omega_{11}(r)+\omega_{13}(r) \{5\cos^{2}(\theta)-1\}]~.
\end{eqnarray}
The fact that $\omega_{0}(r)$ can be analytically solved outside the star is a known result \cite{Hartle2}, and the solution is given as ($r>R$):
\begin{equation}
\label{omega_0out}
\omega_{0}(r)=\frac{B_{1}}{r^{3}}.
\end{equation}
Moreover, similar to the uniformly rotating case, one can also find analytic solutions for $\omega_{11}, ~\omega_{13}$ outside the star. (This is because the equation \eqref{decomposed} can be rewritten outside the star into the form of the hypergeometric equation.)
The analytic solutions are ($r>R$):
\begin{equation}
\label{11}
\omega_{11}(r)=\frac{B_{2}}{r^{3}}
\end{equation}
and 
\begin{widetext}
\begin{eqnarray}\label{13}
\omega_{13}(r)=B_{3}\cdot\left[-\frac{1}{z^{3}}-\frac{5}{z^{2}}-\frac{30}{z}+210-180z+\ln\left\{\frac{z}{z-1}\right\}(120-300z+180z^{2})\right],~~
\end{eqnarray}
\end{widetext}
with $z=r/2M$. ($M$ being the mass of the star.) Although this is maybe not obvious, the solution for $\omega_{13}$ can be shown to behave as $\sim r^{-5}$ when approaching infinity. (All the terms with powers higher than $r^{-5}$ cancel out. The derivation of the solutions and their asymptotic behavior is left for the appendix \ref{frame dragging}.) 
Let us also add that the physical meaning of the constants $B_{1,2}$ is the following: 
\begin{equation}
\frac{B_{1}+\gamma\cdot B_{2}}{2}=J, 
\end{equation}
where $J$ is the angular momentum of the star.

The procedure for numerically computing the frame dragging $\omega$ is the following: First we numerically determine $\omega_0$ by solving eq. (\ref{Hartle}) inside the star for $\gamma = 0$ (the usual Hartle equation from \cite{Hartle2}) with $\omega_0'(0) = 0$ and $\omega_0(0)$ finite. The value $\omega_0(R)$ then fixes the constant $B_1$ from eq. (\ref{omega_0out}), determining the behavior of $\omega_0$ outside the star. The other components $\omega_{11}$ and $\omega_{13}$ are obtained from the numerical integration of eq. (\ref{decomposed}) with $\ell = 1$ and 3, respectively. In order to pick the regular solutions, we write $\omega_{11}$ and $\omega_{13}$ up to the second order in $r$ with regular series expansions near the center as 
\begin{eqnarray}
\omega_{11}(r) = b_0 \left[ 1 + \frac{8\pi}{5}(\epsilon_0+p_0)r^2\right], \nonumber \quad
\omega_{13}(r) = c_0r^2, \nonumber
\end{eqnarray}
and the constants $b_0$, $c_0$ and $B_2$ and $B_3$ (see eqs. (\ref{11}) and (\ref{13})) are determined with a shooting method, by requiring that both $\omega_{11}$ and $\omega_{13}$ and their first derivatives be continuous on the stellar surface. Finally, the total dragging $\omega(r,\theta)$ is computed with eq. (\ref{dragg}).

\begin{figure}[!htb]
\begin{center}
\includegraphics[angle=270,width=1\linewidth]{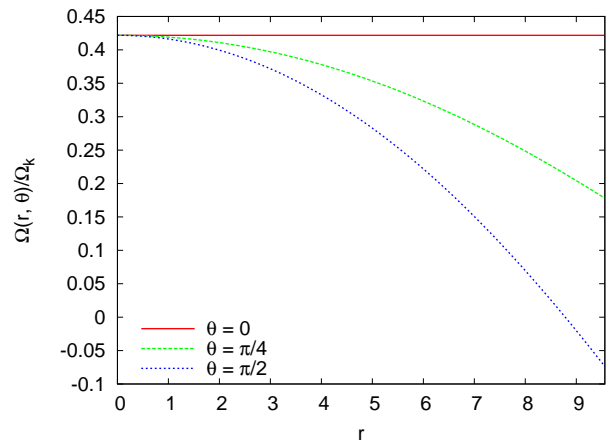}
\end{center}
\caption{Angular velocity $\Omega$ as a function of $r$ for different angles $\theta$. ($\gamma=10^{-2}$.)}
\label{fig:Omega1}
\end{figure}

\begin{figure}[!htb]
\begin{center}
\includegraphics[angle=270,width=1\linewidth]{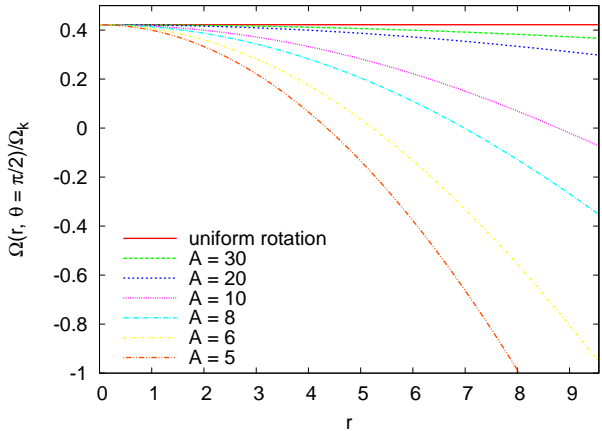}
\end{center}
\caption{Angular velocity $\Omega$ as a function of $r$ for different values of $A=\gamma^{-1/2}$.}
\label{fig:Omega2}
\end{figure}

\begin{figure}[!htb]
\begin{center}
\includegraphics[angle=270,width=1\linewidth]{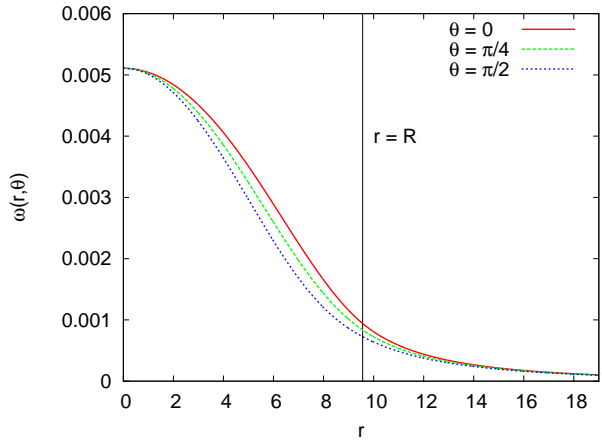}
\end{center}
\caption{Frame dragging function as a function of $r$ for different values of angle $\theta$. ($\gamma=10^{-2}$.)}
\label{fig:omega1}
\end{figure}

\begin{figure}[!htb]
\begin{center}
\includegraphics[angle=270,width=1\linewidth]{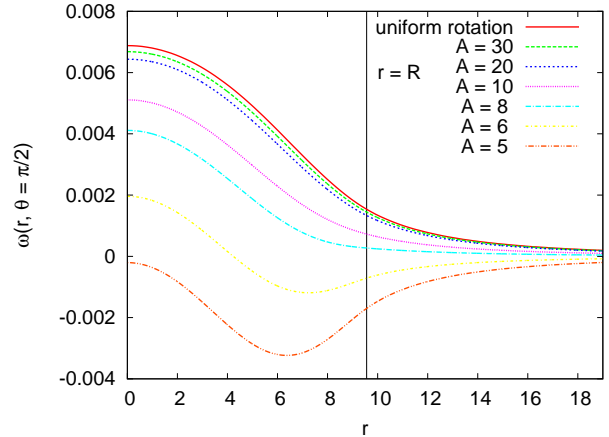}
\end{center}
\caption{Frame dragging function as a function of $r$ for different values of $A=\gamma^{-1/2}$.}
\label{fig:omega2}
\end{figure}

\begin{figure}[!htb]
\begin{center}
\includegraphics[angle=270,width=1\linewidth]{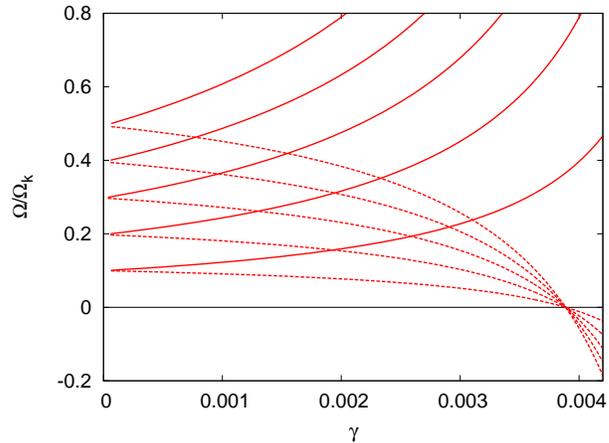}
\end{center}
\caption{The solid lines represent the central angular velocities and the dashed lines the angular velocities at the surface as functions of $\gamma$, for different values of angular momenta. We use the star with the compactnes $M/R=0.1$ and the equation of state with $N=1.5$, $K=10.86$. (The total mass of the star is $M=1.47$.) }
\label{fig:gamma1}
\end{figure}

\begin{figure}[!htb]
\begin{center}
\includegraphics[angle=270,width=1\linewidth]{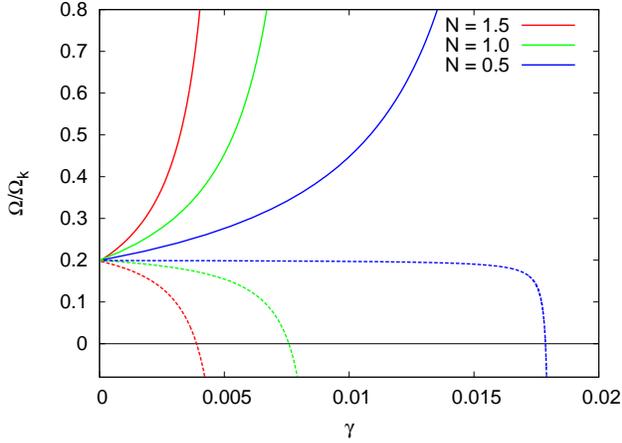}
\end{center}
\caption{The solid lines represent the central angular velocity and the dashed lines the angular velocity at the surface as functions of $\gamma$, for different stars with different angular momenta. The mass of the star decreases for different lines representing different cases from left to right. The compactness of the star is in all the three cases $M/R=0.1$ and the parameters of the equation of state are (from right to left in the plot) $N=0.5,~ 1,~ 1.5$,~ $K=78106,~ 100,~ 10.86$.  }
\label{fig:gamma2}
\end{figure}

\begin{figure}[!htb]
\begin{center}
\includegraphics[angle=270,width=1\linewidth]{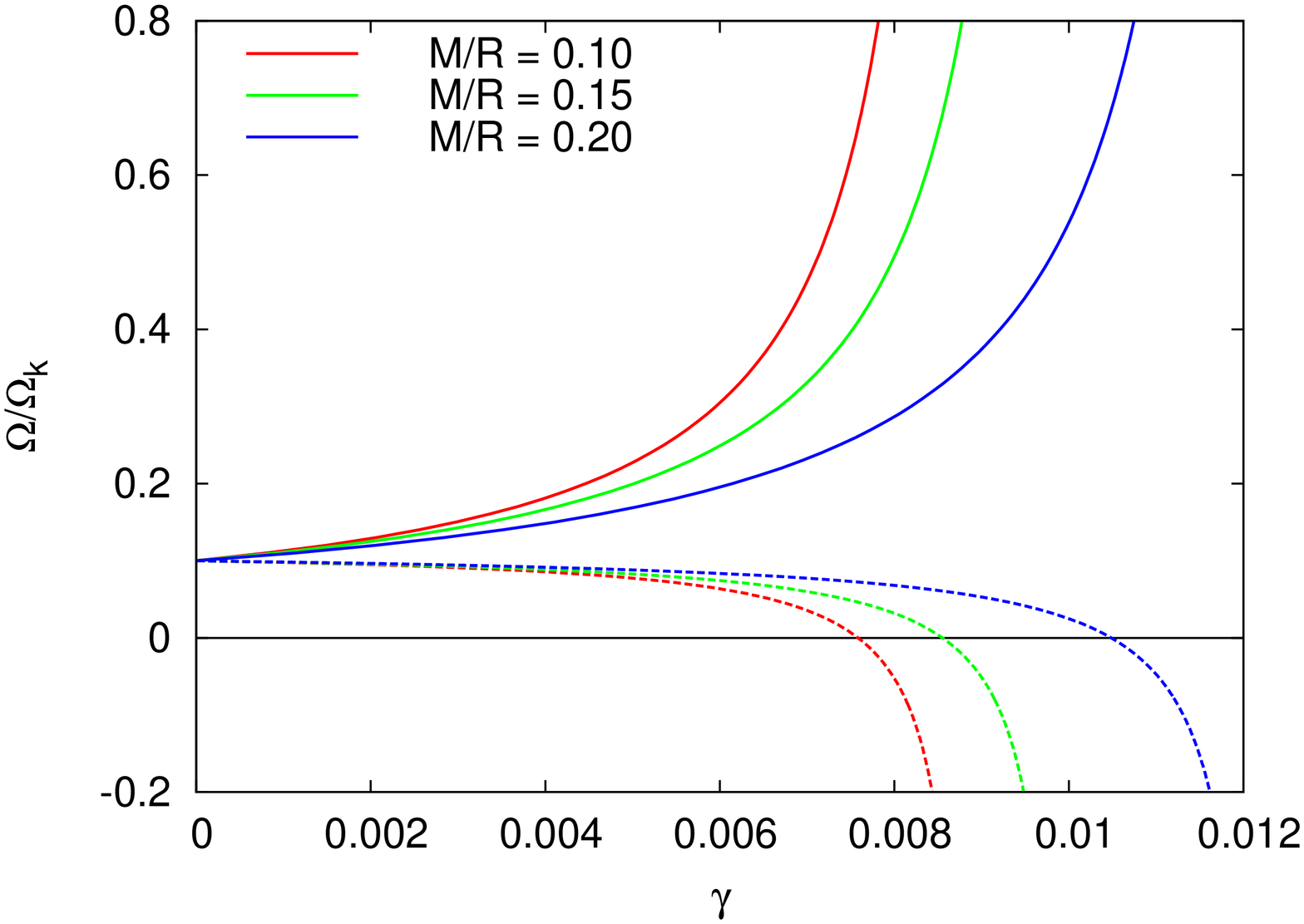}
\end{center}
\caption{The solid line represents the central angular velocity and the dashed line the angular velocity at the surface as functions of $\gamma$, for different stars with different angular momenta. The compactness of the stars grows for different lines representing different cases from left to right and takes the values $M/R=0.1~,0.15,~0.2$.~The star has the equation of state with $N=1$ and $K=100$. (The corresponding stellar masses are $M=1.06,~1.4,~1.62$.)  }
\label{fig:gamma3}
\end{figure}

In the figures \ref{fig:Omega1} and \ref{fig:Omega2},  we present some plots of the angular velocity $\Omega$ as a function of the radial coordinate, for different values of angle $\theta$ and different values of the inverted $\gamma$ parameter. We also present (figures \ref{fig:omega1} and \ref{fig:omega2}) plots of the frame dragging $\omega$ as a function of the radial coordinate, for different values of angle $\theta$ and different values of the inverted $\gamma$ parameter. (The star is taken in the units $c=G=M_{\odot}=1$ with the compactness $M/R=0.15$ and with the equation of state parameters $N=1$, $K=100$. This gives the stellar mass to be $M=1.4$. Note also that unless explicitly stated otherwise, we will use everywhere in the paper the units $c=G=M_{\odot}=1$.)
In the figures \ref{fig:gamma1}, \ref{fig:gamma2}, \ref{fig:gamma3} we show the dependence of the central and the surface angular velocities on the $\gamma$ parameter. (All the plots are taken at the equatorial plane.) The angular velocities are normalized by the Keplerian mass shedding limit, $\Omega_K$. The minimal bounds on the $\gamma$ parameter are given by the equilibrium model, when either the central angular velocity reaches the value $\sim 0.8 \Omega_K$, or when the surface angular velocity reaches zero. In this sense the minimal bounds on $\gamma$ are obtained naturally in the equatorial plane, as one can easily analytically observe that the second bound on $\gamma$, given by the surface angular velocity, has lowest value in the equatorial plane.  (The first bound given by the central angular velocity is independent on $\theta$.) 
In the next section, after computing the f-modes we  further restrict the value of $\gamma$ by comparing our results for the f-mode frequencies with the results of \cite{Passamonti2}. We confirm there is a very good agreement (less than 3 \% error) up to the value $\gamma \sim 30^{-2}$, but for $\gamma \sim 20^{-2}$ the error is already 25 \%, so the bound ($\gamma_{B}$) on $\gamma$ can be put as $\gamma_{B}\lesssim 20^{-2}$.

\section{Linearized perturbation equations for the fluid} 
\label{sec:perturb}

We
work in the Cowling approximation, thus we have only fluid perturbation
variables, in particular: ~$\delta \epsilon,~ \delta p,~ \delta
u^{\mu}$.~ There are two more principles one uses to reduce the number of the variables to four: the four-velocity normalization condition $\delta (u^{\mu}u_{\mu})=0$, and the fact that the perturbed fluid is barotropic:
\begin{equation}\label{first}
\delta\epsilon=\frac{\epsilon+p}{\Gamma p}~\delta p.
\end{equation}
The remaining variables are: $\delta u^{r}, \delta u^{\theta}, \delta u^{\phi}, \delta Q$, with  ~$\delta Q=\delta p/(p+\epsilon)$. The dynamical equations are obtained from the three independent components of the perturbed Euler equation $\delta((\delta^{\mu}_{\kappa} + u^{\mu}u_{\kappa})T^{\kappa\nu}_{;\nu})=0$ and the perturbed energy conservation equation $\delta(u_{\kappa}T^{\kappa\nu}_{;\nu})=0$.  The final four equations for the linearized dynamics of the fluid can be written as: 
\begin{widetext}
\begin{eqnarray}\label{1}
\delta u^{\theta}_{,t}+\Omega\cdot\delta u^{\theta}_{,\phi}+\sin^{2}(\theta)\cdot\left[\omega_{,\theta}-2\cot(\theta)\cdot(\Omega-\omega)\right]\cdot\delta u^{\phi}=-\frac{e^{\nu/2}}{r^{2}}\cdot\delta Q_{,\theta}~,~~
\end{eqnarray}

\begin{eqnarray}\label{2}
\delta u^{\phi}_{,t}+\Omega\cdot\delta u^{\phi}_{,\phi}
+\left[(\Omega-\omega)_{,r}+\left(\frac{2}{r}-\nu_{,r}\right)\cdot(\Omega-\omega)\right]\delta u^{r}+~~~~~~~~~~~~~~~~~~~~~~~~~~~~~~~~~~~~~~~~~~~~~~~~~~~~~~~~~\nonumber\\
\left[(\Omega-\omega)_{,\theta}+2\cot(\theta)\cdot(\Omega-\omega)\right]\cdot\delta
u^{\theta}
=-\left[\frac{e^{\nu/2}}{r^{2}\sin^{2}(\theta)}\cdot\delta
Q_{,\phi}+e^{-\nu/2}(\Omega-\omega)\cdot\delta
Q_{,t}\right],~~~~~~~~~~~~~~
\end{eqnarray}

\begin{eqnarray}\label{3}
e^{\lambda}\cdot\delta u^{r}_{,t}+e^{\lambda}\cdot\Omega\cdot\delta u^{r}_{,\phi}+r\cdot\sin^{2}(\theta)\cdot\left[r\cdot\omega_{,r}+(r\cdot\nu_{,r}-2)\cdot(\Omega-\omega)\right]\cdot\delta u^{\phi}=-e^{\nu/2}\cdot\delta
Q_{,r}~, ~~~~
\end{eqnarray}

\begin{eqnarray}\label{4}
\delta Q_{,t}+\Omega\cdot\delta Q_{,\phi}+e^{\nu/2}\cdot\frac{\Gamma p}{\epsilon+p}\left[e^{-\nu}r^{2}\sin^{2}(\theta)\cdot(\Omega-\omega)\cdot\delta u^{\phi}_{,t}+\delta u^{r}_{,r}+\delta u^{\theta}_{,\theta}+\delta u^{\phi}_{,\phi}\right]=~~~~~~~~~~~~~~~~~~~~~\nonumber\\
=-e^{\nu/2}\cdot\left\{\frac{\Gamma p}{\epsilon+p}\left[\frac{\nu_{,r}+\lambda_{,r}}{2}+\frac{2}{r}\right]-\frac{\nu_{,r}}{2}\right\}\cdot\delta u^{r}-e^{\nu/2}\cdot\frac{\Gamma p}{\epsilon+p}\cdot\cot(\theta)\cdot\delta u^{\theta}~.~~~~~~~~
\end{eqnarray}
\end{widetext}

\section{Numerical results for the modes}
\label{sec:numerics}

We used a 2D Lax-Wandroff scheme for solving the perturbation equations and the frequencies for the fluid modes were obtained through the Fourier transform of the time evolution of $\delta p$ at a given point inside the star. For the numerical integration we used the form of the equations \eqref{1}, \eqref{2}, \eqref{3}, \eqref{4} rewritten in the variables ~$\{\delta p, ~f^{i}\}$,~ where $f^{i}$ is a momentum-like variable defined as~ $f^{i}=(p+\epsilon)\delta u^{i}$ (similarly to what was done in \cite{Andersson} for Newtonian polytropes). We used symmetric boundary conditions for $\delta p$ (at the equatorial plane) for the f-modes and antisymetric boundary conditions for $\delta p$ for the r-modes. Also we used the regularity condition at both the radial center and the rotational axis. For the r-modes we used the initial data from \cite{Owen} and for the f-modes we used the initial value conditions from  \cite{Andersson}.  

\begin{figure}[!htb]
\begin{center}
\includegraphics[angle=270,width=1\linewidth]{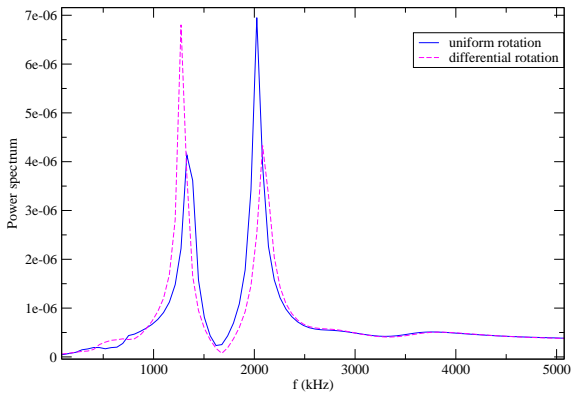}
\end{center}
\caption{The power spectrum obtained from our time evolution data for two stars with the same angular momentum $J$, with uniform (solid line) and differential rotation with $\gamma = 0.003$ (dashed line). Both stars have $J = 0.2$ and equation of state with N = 1.5, K = 10.86 (compactness M/R = 0.14 and mass M = 1.5).}
\label{fig:DFT}
\end{figure}

In figure \ref{fig:DFT} we have a representative power spectrum obtained from our evolution data. One can see in this plot the correction to the rotational split added by the differential rotation. The relative heights of the peaks are rather arbitrary, and depend only on the initial data used. The width of the peaks is caused by the numerical dissipation of the algorithm used: combining that and our comparisons with values from the literature, we estimate that our numerical error in the determination of the frequencies (see below) is within 3 \%.

We present in figure \ref{fig:fmodes1} detailed results for the f-modes in the appropriate range of validity of $\gamma$. (As previously mentioned, for $\gamma<\gamma_{B}\lesssim 20^{-2}$ the agreement with the results of \cite{Passamonti2} is within less than 3 \% error.) We present in this figure an equivalent to the correction to the frequency given by the rotational splitting of the f-modes in the uniformly rotating case, normalized now by the surface angular velocity of the star at the equatorial plane, $\Omega_e$, in units of $\Omega_K$. The two different data sequences correspond to constant $J$ sequences of two polytropic stars with different polytropic indexes $N$, but approximately the same compactness. We can see that, for larger value of $N$, the correction starts with lower values, but grows faster with increasing differential rotation. This very fast growth shows a limitation of our first order treatment of the differential rotation. As seen in \cite{Passamonti2}, when second order terms are takend into account, this growth becomes much less steep and much more "well-behaved".

We also present in tables \ref{tab:rmodes1} and \ref{tab:rmodes2} results for the r-mode frequencies, which were not computed in \cite{Passamonti2}.  We computed r-modes also for the uniformly rotating case, compared our results with \cite{Ruoff} and they agreed again with less than 3 \% error. In \cite{Ruoff} they used decomposition of the perturbations into spherical harmonics and after truncating the coupled equations at $\ell_{max}$, they time evolved the 1D wave function to obtain the frequencies. Table \ref{tab:rmodes1} presents the values of the r-modes frequencies for a sequence of stars with constant angular velocity at the center $\Omega_c$ and increasing differential rotation, while table \ref{tab:rmodes2}  presents the same results for a sequence of stars with constant angular momentum $J$.

\begin{figure}[!htb]
\begin{center}
\includegraphics[angle=270,width=1\linewidth]{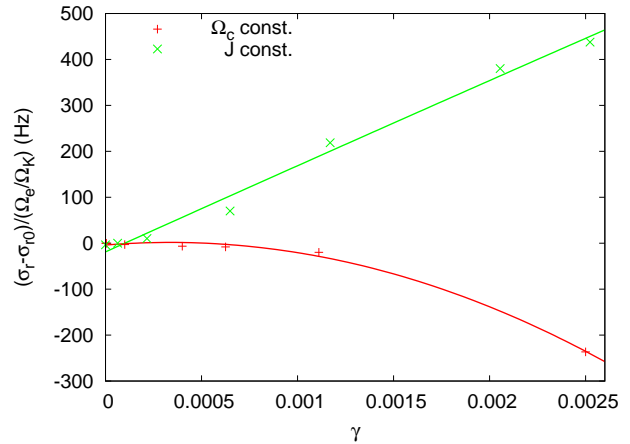}
\end{center}
\caption{The equivalent of the rotational correction for the r-modes for sequences with constant $J$ and sequences with constant angular velocity at the center $\Omega_{c}$. This are the values from tables \ref{tab:rmodes1}, \ref{tab:rmodes2}, so the star is taken with $N = 1$, $K = 100$, $M/R = 0.15$, (and $M = 1.4$).}
\label{fig:rmodes1}
\end{figure}

In the figure \ref{fig:rmodes1} we present an equivalent to the correction given in figure \ref{fig:fmodes1}, but using now as central value the r-mode frequency for a uniformly rotating star, $\sigma_{r0}$. We used the results from both tables \ref{tab:rmodes1} and \ref{tab:rmodes2} in order to calculate these results. Note that the effect of the differential rotation seems to be much weaker for the r-modes than for the f-modes (the scale of the vertical axis is now in Hz, and not in kHz as it was in figure \ref{fig:fmodes1}).

\begin{table}[h]
\centering
\begin{tabular}{|c| c| c|c| }
\hline
A  &   $\sigma_{r}$(Hz) &    $\Omega_{e}$(Hz)  & J \\	\hline
1000 &   491.55 &       347	& 0.2135 \\  \hline
500 &    491.49  &	347  &	0.2134	 \\   \hline
100  &  490.78	 &    	343 &	0.2114	 \\  \hline
50  &   489.92	 &     	331 &	0.2050 \\  \hline	
40   &  489.57	 &    	321 &	0.2002 \\  \hline	
30   &  486.92	 &    	302 &	0.1898 \\  \hline	
20  &    446.90	  &    	245 &	0.1603	 \\  \hline
\end{tabular}
\caption{The table of r-mode frequencies $\sigma_{r}$ with constant angular velocity at the center ($\Omega_{c}$) and changing $A=\gamma^{-1/2}$. ($\Omega_{e}$ is surface angular velocity at the equatorial plane.) The star is taken with $N = 1,~ K = 100,~  M/R = 0.15,~ \Omega_{c}=347$ Hz,~ $ M = 1.4$.}
\label{tab:rmodes1}
\end{table}

\begin{table}[h]
\centering
\begin{tabular}{|c| c| c|c| }
\hline
 A &    	$\sigma_{r}$(Hz) &   $\Omega_{c}$(Hz) &	$\Omega_{e}$(Hz) \\ \hline
2262 &	  491.55 &	      	347 &	347 \\	\hline
126.14 & 492.47 &	   	349 &	346 \\ \hline	
68.03 &  	495.19 & 	354 &	346 \\  \hline
39.26 &	511.02 &	  371 &	 343  \\ \hline	
29.24	& 549.59 &	     	393 &	339 \\ \hline	
22.06 &	589.43 &		436 &	331 \\ \hline	
19.91	& 602.76 &	     	463 &	327 \\ \hline	
\end{tabular}
\caption{The table of r-mode frequencies $\sigma_{r}$ with constant $J$ and changing $A=\gamma^{-1/2}$. The star is taken with $N = 1,~ K = 100,~  M/R = 0.15,~ J=0.2135,~  M = 1.4$.}
\label{tab:rmodes2}
\end{table}

\begin{figure}[!htb]
\begin{center}
\includegraphics[angle=270,width=1\linewidth]{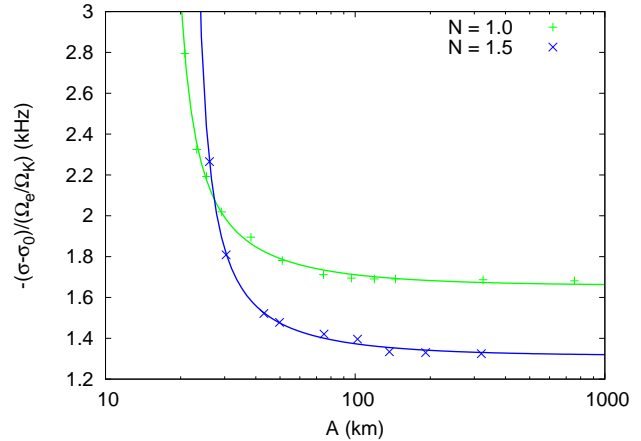}
\end{center}
\caption{The rotational correction for the $f_{-}$-mode frequencies within the range of reliability of $A=\gamma^{-1/2}$ for the stars with the equation of state with N=1, 1.5, K=100, 10.86, and with compactness M/R=0.15, 0.14, (and masses M= 1.4, 1.5) in sequences with constant $J=0.2$. The quantity plotted is analogous to the quantities defined in \cite{Kojima} for the uniformly rotating case, with $\sigma$ the frequency of the $f_-$ for the given value of $A$, and $\sigma_0$ the frequency of $f_-$ for the correspondent non-rotating star.}
\label{fig:fmodes1}
\end{figure}

\section{An effect of differential rotation on gravitational wave
 emission from the f-mode}
\label{sec:gw}

We here study an effect of differential rotation on gravitational radiation
by using a simple analysis. We use the Newtonian mass quadrupole formula 
to evaluate the gravitational wave emission. Luminosity of an eigenmode is
computed by using the eigenfunction extracted by DFT. The luminosity is
the quadratic functional of the eigenfunction. On the other hand, we
compute the kinetic energy of the eigenmode, which is also a quadratic
functional of the eigenfunction. Taking the ratio of the luminosity and
the energy, we obtain an inverse of the damping timescale of the eigenmode due to
gravitational radiation. By comparing the timescale for different degrees
of differential rotation, we evaluate how differential rotation affects
gravitational emission from the eigenmode oscillation.

\subsection{Extracting eigenfunctions}
\begin{figure}[!htb]
  \includegraphics[width=0.45\linewidth]{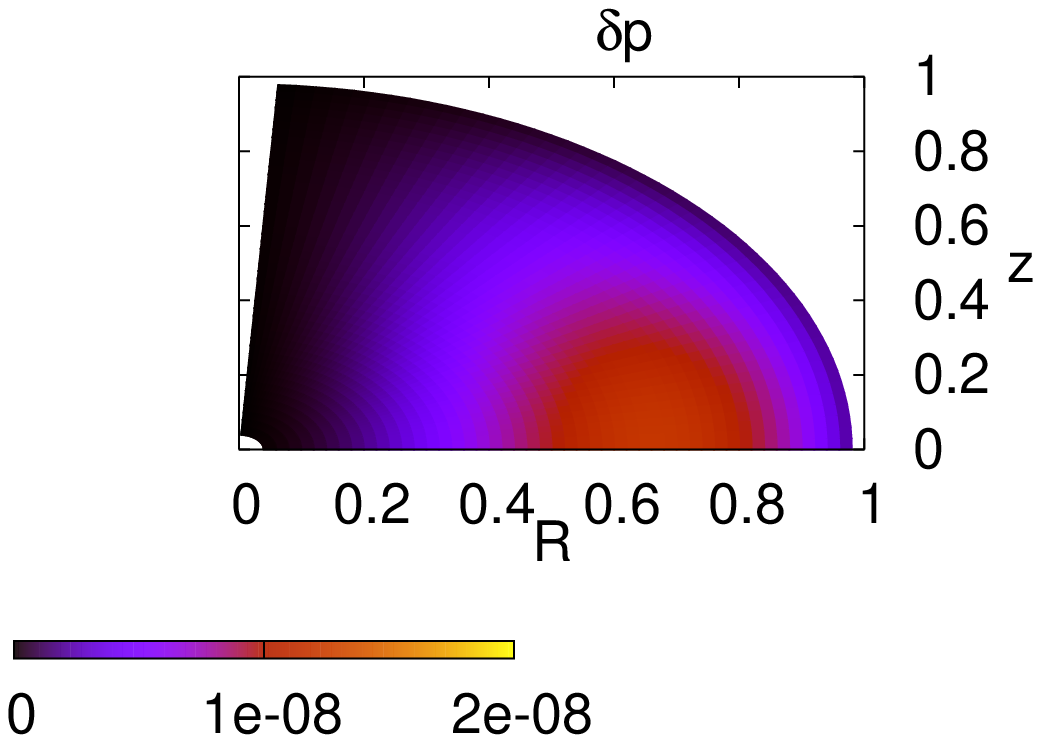}
   \includegraphics[width=0.45\linewidth]{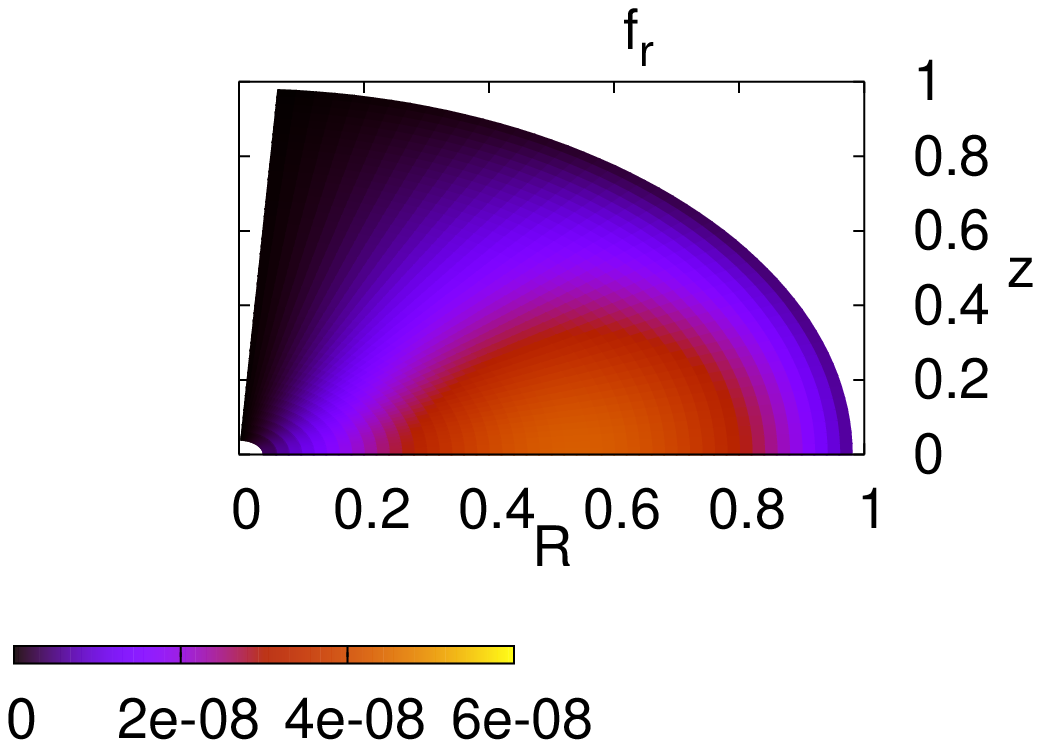}
    \includegraphics[width=0.45\linewidth]{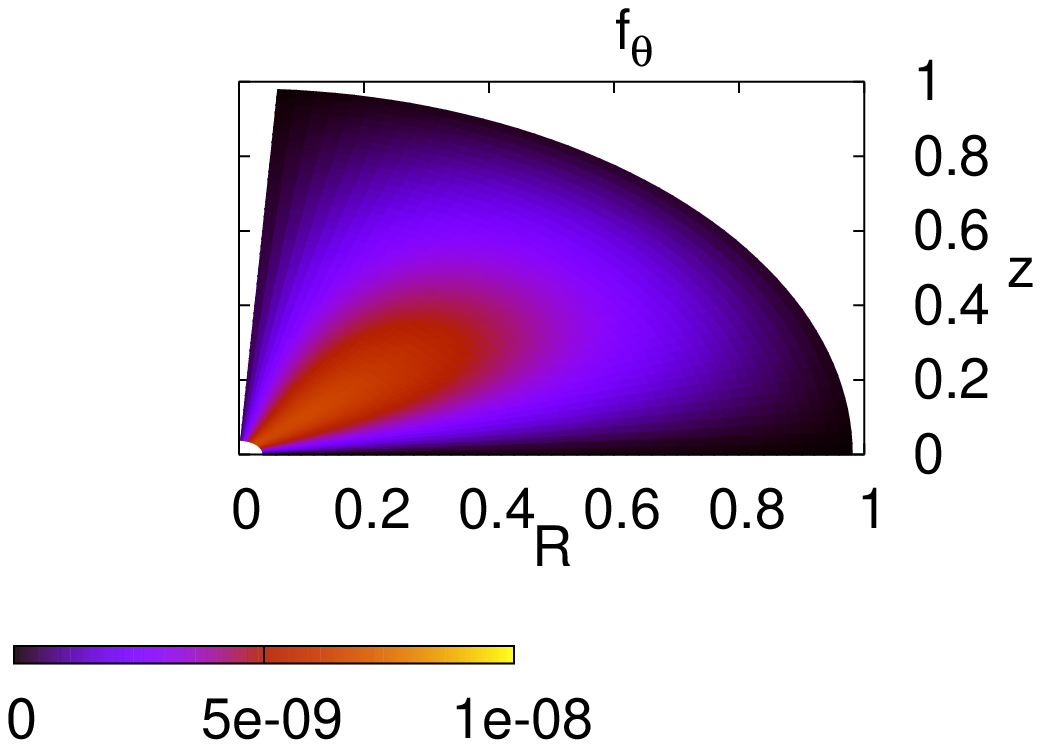}
     \includegraphics[width=0.45\linewidth]{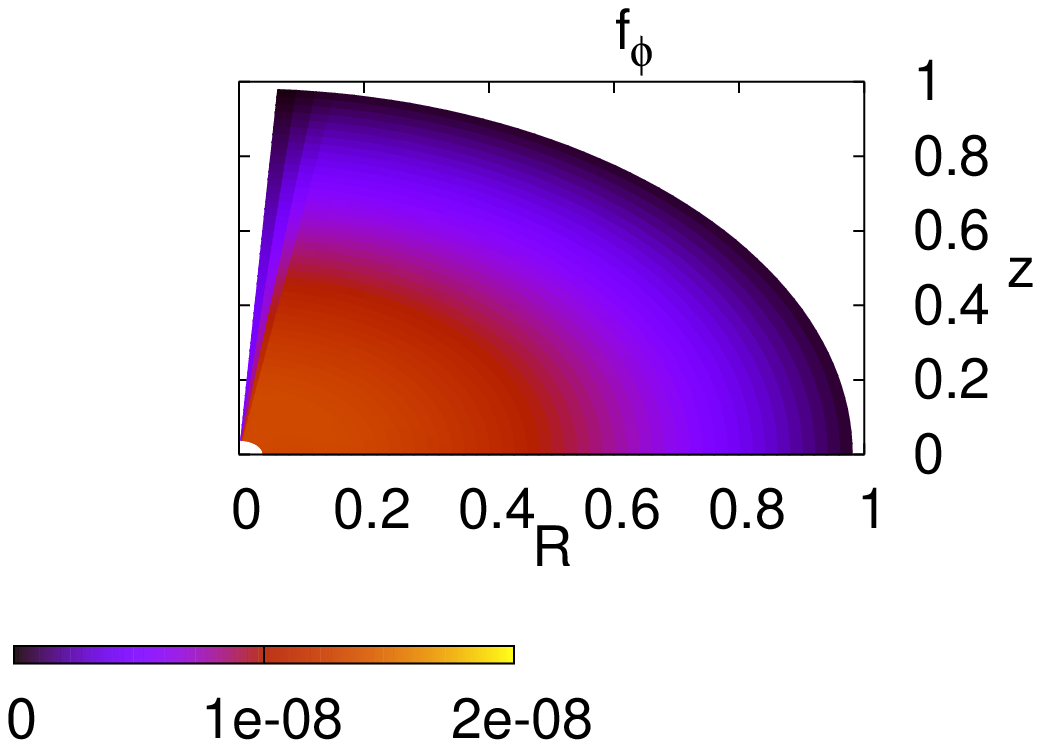}
 \caption{Eigenfunction of $\ell=|m|=2$ $f_{+}$ mode for a differentially
 rotating star with $A=25.89$. The upper half of the meridional section
of the star is shown, with $R$ the equatorial coordinate distance and $z$
the coordinate distance parallel to the rotational axis. Up-Left: $\delta p$. Up-right: $f_r$. Down-left:
 $f_\theta$. Down-right: $f_\phi$. The coordinate distance is normalized by the stellar radius.
}
\label{fig:eigen}
\end{figure}

We extract the eigenfunction of the f-modes using the procedure described as follows. On each spatial grid point, we performed a DFT of the physical variables to extract their power spectra.
An eigenmode corresponds to a peak in the spectrum whose frequency
is constant in space and is shared by different physical variables.
We approximate the peak
with a Lorentzian profile and extract the central frequency, the peak
amplitude, and its width. The collection of the amplitudes on each
grid point gives the absolute value of the eigenmode excited in the simulation.
We obtained sufficiently smooth eigenfunction profiles for the f-modes,
but we failed to extract higher order p-modes. It might be that we
need to prepare initial data that contains the p-mode component with a 
larger amplitude than our current cases. In figure \ref{fig:eigen} we present some typical results obtained for $\delta p$, $f_r$, $f_{\theta}$ and $f_{\phi}$ for the $f_+$-mode which limits to the $\ell=m=2$ f-mode in the non-rotating limit.

We also obtain the eigenfunctions for the $f_-$-mode, which shows qualitatively very similar
functional profiles. Althogh we tried to extract p-mode eigenfunctions, but it was been hindered
by numerical noise.

\subsection{Radiation timescale}

Formulae for the gravitational radiation from stellar oscillations are
found in \cite{ipser_lindblom91} in the Newtonian limit. The energy of linear perturbation is
defined as Eq.(15) there. Since our model is in the Cowling approximation,
we have no gravitational perturbation (which is $\delta U$ in \cite{ipser_lindblom91}).
Therefore, we approximate the linear perturbation energy as,
\begin{equation}
	\hat{E} = \int \rho~\delta v^a \delta v^*_a dV,
\end{equation}
where we assume the potential energy is regarded as the same size as
the kinetic energy (i.e. we assume "equipartition" as in a simple oscillator).

Energy loss by gravitational radiation is given by
\begin{widetext}
\begin{equation}
\frac{dE}{dt} = -\sum_{\ell\ge m}
(-1)^\ell N_\ell ~\Re\left[
\frac{d^{2\ell+1}}{dt^{2\ell+1}}\delta D_\ell^m \cdot
\left(\frac{d}{dt}\delta D_\ell^{*m}-im\Omega\delta D_\ell^{*m}
\right)
\right]
\end{equation}
\end{widetext}
where
\begin{equation}
	N_\ell = \frac{4\pi G}{c^{2\ell+1}}\frac{(\ell+1)(\ell+2)}
	{\ell(\ell-1)[(2\ell+1)!!]^2},
\end{equation}
and
\begin{equation}
	\delta D_\ell^m = \int \delta\rho ~ r^\ell Y_\ell^{*m} dV.
\end{equation}

Since the amplitude of the eigenmodes extracted from the linear evolution
is arbitrary, we should study the damping timescale $\tau_g$ to characterize
the efficiency of gravitational radiation,
\begin{equation}
	\tau_g^{-1} = \frac{1}{\hat{E}}\frac{dE}{dt}
\end{equation}
instead of an absolute amount of energy radiated.

To compute $\tau_g$ numerically, we need to define and compute the perturbed quantities
appearing in the equations above. We need to have 
$\delta v^i$ (perturbed 3-velocity), $\delta\rho$ (perturbed mass density)
and $dV$ (3-volume element). As for the density we used the perturbed rest mass density.
The volume element is defined in the spatial hypersurface with $t=$const., where $t$ is Schwarzschild
time. The 3-metric $\gamma_{ij}$ is naturally chosen 
as $\gamma = {\rm diag}(e^{2\lambda}, r^2, r^2\sin^2\theta)$. Thus the corresponding
volume element is $dV = e^\lambda r^2\sin\theta dr d\theta d\varphi$.

As for the 3-velocity perturbation, we adopted the definition below.
The 3-velocity perturbation is expressed by the perturbed 4-velocity components as
\begin{equation}
	\delta v^i = \delta\left(\frac{u^i}{u^t}\right)
	= \frac{\delta u^i}{u^t} - \frac{u^i}{(u^t)^2} \delta u^t,
\end{equation}
where $u^\mu$ is the 4-velocity.
Then the components of the perturbed velocity (in coordinate basis) are expressed by
our basic variables $f_i~(i=r,\theta,\varphi)$ as
\begin{equation}
	\delta v^r = \frac{\delta u^r}{u^t} = \frac{f^r}{(\epsilon+p) u^t},
\end{equation}
\begin{equation}
	\delta v^\theta = \frac{\delta u^\theta}{u^t} = \frac{f^\theta}{(\epsilon+p) u^t},
\end{equation}
\begin{widetext}
\begin{equation}
	\delta v^\varphi 
	= \frac{\delta u^\varphi}{u^t}-\frac{u^\varphi}{(u^t)^2}\delta u^t 
	= \left[1+\Omega(\Omega-\omega)r^2\sin^2\theta
	 \right]^{-1}
	\frac{f^\varphi}{(\epsilon+p) u^t},
\end{equation}
\end{widetext}
where $f^i~ (r,\theta,\varphi)$ are defined as before as 
$f^i = (\epsilon+p)\delta u^i$. We have used here
\begin{equation}
	u^t = [e^{2\nu}-r^2\sin^2\theta (\Omega-\omega)^2]^{-\frac{1}{2}},
\end{equation}
and
\begin{equation}
	\delta u^t = \frac{1}{u^t}
	\frac{r^2\sin^2\theta (\Omega-\omega)}
	{e^{2\nu}-r^2\sin^2\theta (\Omega-\omega)^2} \delta v^\varphi.
\end{equation}

Together with the equilibrium values of $\rho$, $\Omega$ and metric coefficients,
these perturbed variables are used to compute $\hat{E}$, $dE/dt$ and $\tau_{GW}$.

\subsection{Results}

\begin{figure}
 \includegraphics[width=1\linewidth]{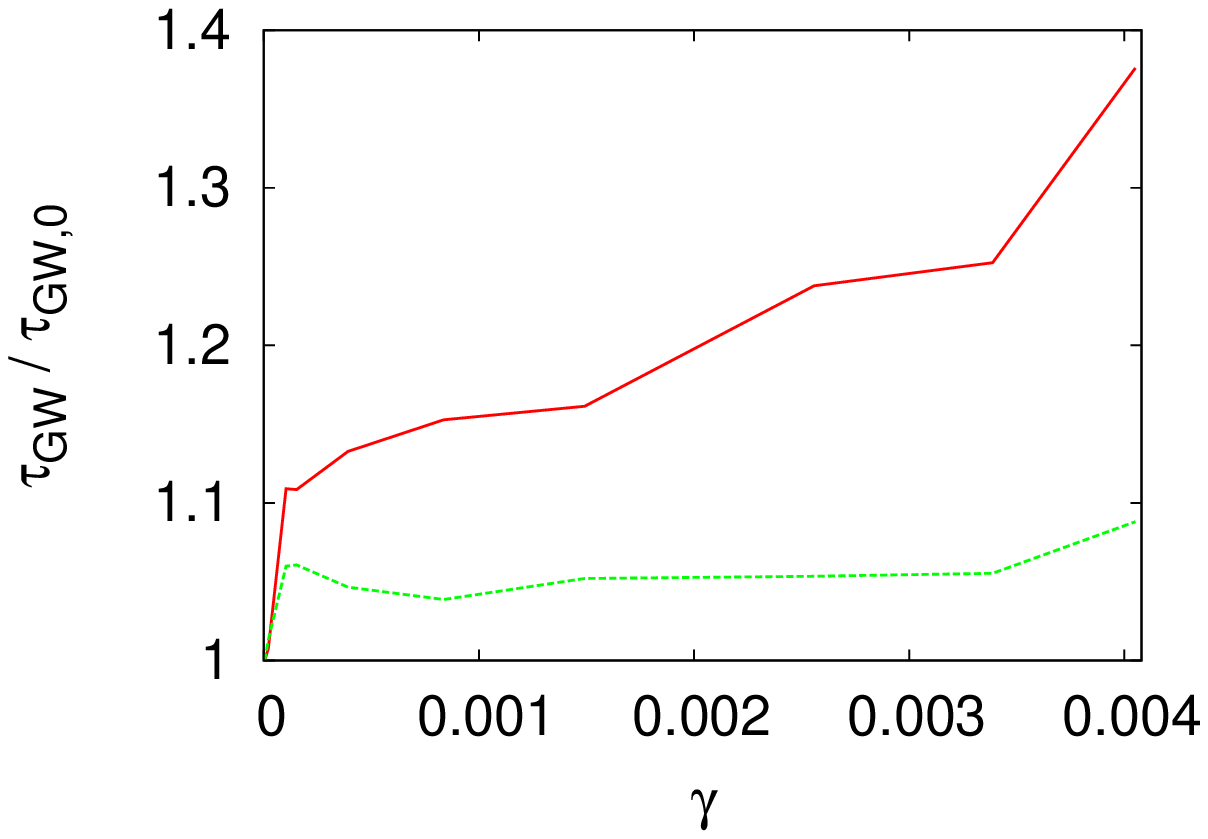}
 \caption{Damping timescale of the eigenmode due to gravitational radiation,
 $\tau_{GW}$, for $\ell=|m|=2$ f-modes. The sequences are obtained by
 fixing the angular momentum and
 increasing $\gamma$, the degree of differential rotation. The timescale
 is normalized by that of the same eigenmode
 in a uniformly rotating star with the same angular momentum. The solid curve
 is for the $f_{-}$ mode (counter-rotating mode) and the dashed one is for
 the $f_{+}$ mode (prograde mode).}
 \label{fig:tauGW}
\end{figure}

We compare the gravitational damping timescale $\tau_{GW}$ for different
degrees of differential rotation parametrized by $\gamma$ for a fixed
value of total angular momentum of the equilibrium star. In figure \ref{fig:tauGW}
we plot sequences of $\tau_{GW}$ for the $\ell=|m|=2$ f-modes with the
angular momentum $J=0.2$.

In figure \ref{fig:tauGW}, $\tau_{GW}$ is normalized by the corresponding
timescale for a uniformly rotating star $\tau_{GW,0}$. $\gamma=0$
corresponds to the uniformly rotating model. A larger value of
$\tau_{GW}$ means smaller amount of gravitational radiation from the
eigenmode. We see that the emissivity of gravitational radiation from each
mode is reduced by introducing differential rotation. For the
counter-rotating f-mode, this may be partly because the eigenfrequency is
decreasing as we increase the degree of differential rotation. However
it does not explain the increase of $\tau_{GW}$ for the prograde f-mode,
whose frequency increases as we increase the degree of differential
rotation. For the $f_+$ mode with differential rotation, the emissivity enhancement due to the increase of the frequency
may be canceled by a modification of the eigenfunction from that of the uniformly
rotating case which reduces mass multipoles.

\section{Conclusions}
\label{sec:conclusions}

In this work we dealt with a slowly rotating relativistic polytrope, such that has a nearly uniform rotation profile. We generalized the old result of \cite{Hartle2}, (for the equilibrium model), for the first order deviations from the uniform rotation. Similar to \cite{Hartle2} we are able to also provide an analytical solution for the metric in the exterior of the star. Furthermore, we used our equilibrium model to numerically compute (in the Cowling approximation) both f and r-mode frequencies. We also estimated the range of validity of our first order approach in the differential rotation parameter $\gamma$: We used the consistency conditions of the equilibrium model to constrain the domain of the $\gamma$ parameter, and some further restrictions were obtained by comparing our results for the f-modes with the known results in the literature. We provided detailed plots of the f-modes for different polytropes with different compactness/angular momenta and also provided some new results for the r-mode frequencies. 
By using a DFT we extracted the low order f-mode eigencfunctions from the evolution data. With the eignfrequencies and their eigenfunctions, the damping time of the oscillation due to gravitational radiation was estimated. Along the stellar models with a constant value of total angular momentum, we see a larger damping time as we increase the $\gamma$ parameter to characterize the degree of differential rotation. This suggests that the inclusion of the differential rotation with our functional form tends to suppress the emission of gravitational wave
for f-modes.

\medskip

{\bf Acknowledgments:}  This research was supported by FAPESP and the Max Planck Society. SY thanks the Center for Mathematics, Computation and Cognition at UFABC for the financial support on his stay at UFABC. The authors wish to thank Luciano Rezzolla for useful discussions on r-modes and invaluable help on the development of the time evolution code.
\appendix

\section{The equilibrium model}

\subsection{Analysis of the regularity of solutions}\label{regularity}

Let us show that if $C_{\ell}=0$, ~$\ell>1$,~ there does not exist an everywhere regular solution. In such case we are looking for a solution of the equation:
\begin{eqnarray}\label{decomposed2}
[r^{4}j~\omega'_{1 \ell}]'=~~~~~~~~~~~~~~~~~~~~~~~~~~~~~~~~~~~~~~~~~~~~\\
e^{\lambda}~j~r^{2}\left[\ell(\ell+1)-2+16\pi r^{2}(\epsilon+p)\right]\omega_{1\ell}.\nonumber
\end{eqnarray}
However, if a solution goes to zero at $r\to 0$ and in the same time at $r\to\infty$ it must have at some point a local maximum/local minimum depending on whether it is approaching the zero at infinity from the negative values (close to the infinity the solution is negative), or from the positive values (close to the infinity the solution is positive). In case $\omega_{1\ell}$ is close to infinity positive it must have a local maximum where $\omega_{1\ell}$ is positive, in case $\omega_{1\ell}$ is close to infinity negative, it must have somewhere a local minimum where $\omega_{1\ell}$ is negative. Take the equation \eqref{decomposed2} at the point of such a local minimum/maximum (call it $r_{e}$). Then, due to the fact that the first derivative of $\omega_{1\ell}$ vanishes at $r_{e}$ the equation \eqref{decomposed2} can be written as:
\begin{eqnarray}\label{decomposed3}
r_{e}^{4}j~[\omega''_{1 \ell}]_{r_{e}}=~~~~~~~~~~~~~~~~~~~~~~~~~~~~~~~~~~~~~~~~~~~~~~~~~~~\\
e^{\lambda}~j~r_{e}^{2}\left[\ell(\ell+1)-2+16\pi r_{e}^{2}(\epsilon+p)\right]\omega_{1\ell}.\nonumber
\end{eqnarray}
But this equation cannot be fulfilled for a very simple reason: In case of a local maximum $\{\omega''_{1\ell}\}_{r_{e}}<0$, and thus the left side of the equation must be negative, whereas the right side of the equation has always the same sign as $\omega_{1\ell}$, and $\omega_{1\ell}$ is at the local maximum positive. (As we said before the local maximum is taken to be such that $\omega_{1\ell}$ is at the local maximum positive.) In case of local minimum the opposite holds: $\{\omega''_{1\ell}\}_{r_{e}}>0$ and thus the left side of the equation must be positive, whereas the right side of the equation must be negative, because $\omega_{1\ell}$ is at the local minimum negative.
This argumentation means that $\omega_{1\ell}$ for any $\ell\neq 1,3$ must be represented by a trivial, zero solution. But what about ~$\omega_{1 1}$~ and ~$\omega_{1 3}$? (With $\omega_{1 1}$ regularity is not an issue, but one still requires that it shall have zero at infinity.) Here the situation is very different, since on the right side of the equation \eqref{decomposed3} appears another term proportional to $C_{3}$ and this prevents us from determining the sign of the right side of the equation at the extremum of $\omega_{1 1}, ~\omega_{1 3}$. This means both ~$\omega_{1 1},~ \omega_{1 3}$~ are in the game, and one has to proceed further in the analysis.

\subsubsection*{Convergence of solutions near zero for $\ell=1$}

Let us analyse the solutions in case $\ell=1$ near zero. For $\ell=1$ the
equation \eqref{decomposed} turns close to 0 to a more complicated
equation:
\begin{equation}
r~\omega''_{11}+4~\omega'_{11}-K.r~\omega_{11}=0
\end{equation}
with $K=16\pi [\epsilon(0)+ p(0)]$. This equation seems not to have analytic solution, but one can verify that one and only one of the solutions is regular at 0 by decomposing $\omega_{11}$ into McLaurin series (or Taylor series at zero)
\[ \omega_{11}=\sum_{n=0}^{\infty}a_{n}r^{n} \]
and after doing this one obtains
\begin{itemize}
\item  $a_{1}=0$,
\item  $a_{n+2}=K\cdot a_{n}/(n^{2}+5n+4)$.
\end{itemize}
This means there is only one solution that can be decomposed close to 0 to McLaurin series and that strongly indicates that the other solution is singular at 0. Thus qualitatively the case $\ell=1$ is the same than other, $\ell>1$ cases.

\subsection{The frame dragging function solutions outside the star}\label{frame dragging}

Take the equation \eqref{decomposed} outside the star. Consider that outside the star holds ($M$ being the mass of the star):
\begin{equation}
e^{\lambda}=\left(1-\frac{2M}{r}\right)^{-1} .
\end{equation}
Then the equation \eqref{decomposed} can be rewritten outside the star in the form
\begin{equation}\label{outside}
r(r-2M)\omega_{1\ell}''+4(r-2M)\omega_{1\ell}'-[\ell(\ell+1)-2]\omega_{1\ell}=0.
\end{equation}
After redefining the variable ~$r\doteq 2M\cdot z$~ and some algebras one can rewrite the equation \eqref{outside} in the following form:
\begin{equation}\label{hypergeom}
z(1-z)\omega_{1\ell}''+4(1-z)\omega_{1\ell}'+[\ell(\ell+1)-2]\omega_{1\ell}=0.
\end{equation}
By `` $'$ '' we mean here a $z$-derivative. Now consider that
\eqref{hypergeom} is a hypergeometric equation with coefficients
that can be chosen as:
\begin{itemize}
\item  $a=2+\ell$,
\item  $b=1-\ell$,
\item  $c=4$.
\end{itemize}
(Note that $a, ~b$ are in fact minus exponents in the asymptotic
formula \eqref{asymp}.) Unfortunately, due to the fact that $c$ is
an integer, \eqref{hypergeom} cannot be solved by a linear
combination of 2-1 type hypergeometric functions (multiplied by
powers of $z$), which is a generic solution for hypergeometric
equation. Although this cannot be done, let us proceed further:  The
only relevant $\ell$ are $\ell=1,3$, for $\ell=1$ we already know
the general solution and this is:
\begin{equation}
D_{1}\cdot z^{-3}+D_{2}.
\end{equation}
For $\ell=3$ the equation \eqref{hypergeom} becomes
\begin{equation}\label{three}
z(1-z)~\omega''_{13}+4(1-z)~\omega'_{13}+10~\omega_{13}=0.
\end{equation}
The software Mathematica found to the equation \eqref{three} the
following analytic solution:
\begin{widetext}
\begin{eqnarray}\label{solution}
\omega_{13}=D_{1}\left(\frac{2}{3}-\frac{5}{3}z+z^{2}\right)+D_{2}\left[-\frac{1}{z^{3}}-\frac{5}{z^{2}}-\frac{30}{z}+210-180z+\ln\left\{\frac{z}{z-1}\right\}(120-300z+180z^{2})\right].
\end{eqnarray}
\end{widetext}
We can see that the first term is the $z^{2}$ divergent term. It is
slightly less obvious that the other term is actually the convergent
part of the solution behaving as $z^{-5}$. One has to substitute
$z=\epsilon^{-1}$ and take the McLaurin series expansion of
\[\ln\left\{\frac{z}{z-1}\right\}=\ln\left\{\frac{1}{1-\epsilon}\right\}=\epsilon+\frac{1}{2}\cdot\epsilon^{2}+\frac{1}{3}\cdot\epsilon^{3}+... .\]
Then if one expresses the second solution term in $\epsilon$:
\begin{widetext}
\begin{eqnarray}
D_{2}\cdot(-\epsilon^{3}-5\epsilon^{2}-30\epsilon+210-180\epsilon^{-1}+\left[\epsilon+\frac{1}{2}\cdot\epsilon^{2}+\frac{1}{3}\cdot\epsilon^{3}+...\right]\cdot\left\{120-300\epsilon^{-1}+180\epsilon^{-2}\right\}),~~~~~~~
\end{eqnarray}
\end{widetext}
one can observe (after some computation) that all the terms up to
the 5-th power cancel. This means $\omega_{13}$ term is outside the
star given as:
\begin{widetext}
\begin{equation*}
\omega_{13}=D_{2}\left[-\frac{1}{z^{3}}-\frac{5}{z^{2}}-\frac{30}{z}+210-180z+\ln\left\{\frac{z}{z-1}\right\}(120-300z+180z^{2})\right].
\end{equation*}
\end{widetext}

\end{document}